\documentclass[sigconf,screen]{acmart}

\usepackage{algorithm}
\usepackage{makecell}
\usepackage{enumitem}
\usepackage{algorithmic}
\usepackage[utf8]{inputenc}
\usepackage{microtype}
\usepackage{multirow}   
\usepackage{array}
\usepackage{balance}

\newcommand{\answerbox}[1]{%
  \par\addvspace{6pt}%
  \begingroup
  \setlength{\fboxrule}{0.7pt}%
  \setlength{\fboxsep}{3pt}%
  \noindent
  \fcolorbox{black}{green!10}{%
    \parbox{\dimexpr\linewidth-2\fboxrule-2\fboxsep\relax}{%
      \setlength{\leftskip}{1pt}%
      \setlength{\rightskip}{1pt}%
      \noindent #1\par
    }%
  }%
  \par
  \endgroup
}

\AtBeginDocument{%
  }

\setcopyright{cc}
\setcctype{by}
\acmDOI{10.1145/3832783.3834505}
\acmISBN{979-8-4007-2882-2/2026/10}

\acmYear{2026}
\copyrightyear{2026}
\acmConference[ASE '26]{Proceedings of the 41st IEEE/ACM International Conference on Automated Software Engineering}{October 12--16, 2026}{Munich, Germany}
\acmBooktitle{Proceedings of the 41st IEEE/ACM International Conference on Automated Software Engineering (ASE '26), October 12--16, 2026, Munich, Germany}
\received{2026-04-30}
\received[accepted]{2026-07-01}

\begin{document}

\title{DepWareTrans: Dependency-Aware Incremental Repository Migration across Co-executable Languages}

\author{Sivajeet Chand}
\correspondingauthor
\orcid{0009-0000-3930-1343}
\affiliation{%
  \institution{TU Munich}
  \city{Munich}
  \country{Germany}
}
\email{sivajeet.chand@tum.de}

\author{Alexander Pretschner}
\orcid{0000-0002-5573-1201}
\affiliation{%
  \institution{TU Munich}
  \city{Munich}
  \country{Germany}
}
\email{alexander.pretschner@tum.de}

\author{Steve Haupt}
\orcid{0009-0005-0477-4649}
\affiliation{%
  \institution{andrena objects}
  \city{Munich}
  \country{Germany}
}
\email{steve.haupt@andrena.de}

\author{Derui Zhu}
\orcid{0000-0002-9552-0097}
\affiliation{%
  \institution{TU Munich}
  \city{Munich}
  \country{Germany}
}
\email{derui.zhu@tum.de}

\author{Sushant Kumar Pandey}
\orcid{0000-0003-1882-2435}
\affiliation{%
  \institution{University of Groningen}
  \city{Groningen}
  \country{Netherlands}
}
\email{s.k.pandey@rug.nl}

\renewcommand{\shortauthors}{Chand et al.}

\begin{abstract}
  Repository-level code translation is critical for modernizing legacy systems, yet existing approaches based on large language models (LLMs) operate at the file level and fail to scale to codebases with complex inter-file dependencies. This limitation is evident in our industrial setting, where we aim to migrate a production repository (STAR) from Java to Kotlin, but file-level approaches produce fragmented results and fail to achieve end-to-end correctness. In this paper, we show that the primary cause of failure at the repository level is dependency inconsistency. Through an empirical study on open-source and industrial systems, we find that most errors arise from unresolved cross-file dependencies that cannot be effectively addressed by iterative feedback alone. We propose a dependency-aware incremental migration framework that elevates the unit of translation from individual files to dependency-consistent batches. Our approach constructs a dependency graph, groups interdependent files, and performs batched translation with iterative compile- and test-driven validation. We evaluate our method on a 51K line of code (LOC) industrial system and multiple repositories across interoperable language pairs (Java–Kotlin, Java–Scala, and C\#–F\#). On the STAR repository, file-level approaches achieve 38.16\% compilation and 9.39\% test success, whereas our approach achieves 100\% compilation and test success across the evaluated settings, converging within a small number of iterations. These results show that dependency-aware batching improves scalability and reliability in repository-level code translation.
\end{abstract}




\maketitle

\section{Introduction}
Migrating software systems across programming languages is a recurring and high-impact challenge in industrial software development. Organizations undertake such migrations to adopt modern language features, improve maintainability, and align with evolving ecosystems. Large-scale efforts reported by industry, including migrations from Java to Kotlin in Android and enterprise systems, demonstrate both the practical importance and engineering difficulty of these transformations~\cite{android2024kotlinfirst,meta2024java2kotlin,KotlinMigrationStudy}. 

A widely adopted strategy in practice is \emph{incremental migration}, where systems are gradually transformed while remaining continuously buildable and testable. This approach is particularly feasible for \emph{interoperable language pairs}, i.e., languages that share a common runtime and can coexist within the same system (e.g., Java--Kotlin on the JVM, Java--Scala, and C\#--F\# on the .NET platform). Interoperability enables mixed-language systems, allowing organizations to migrate large repositories without requiring disruptive, full-system rewrites~\cite{KotlinJavaDeps}. However, while interoperability makes incremental migration feasible, it does not make it easy. Instead, it introduces a distinct challenge: maintaining consistency across interdependent components written in different languages. In practice, migration is often performed at the level of individual files or small code fragments, using IDE-based converters or large language model (LLM)-assisted translation tools~\cite{jetbrains2026kotlinvscode}. Although effective for small or loosely coupled systems, such file-level workflows do not scale to large, dependency-rich repositories. In particular, translating files in isolation frequently breaks dependency relationships across components. This issue is exacerbated in real-world systems with tightly coupled or cyclic dependencies, where interdependent files must evolve consistently. As a result, even when individual translations are locally correct, the system as a whole often fails to compile or pass tests. This creates significant integration overhead and limits the practical effectiveness of existing migration workflows.

Recent research has begun to explore code translation using LLMs~\cite{CodeFuse13B, Jiao2023Evaluation, Pan2024Lost, Yan2023CodeTransOcean, Yin2024Rectifier}. At repository-scale, approaches such as Syzygy~\cite{Shetty2024Syzygy} and Oxidizer~\cite{Zhang2024Scalable} decompose programs into fine-grained units (e.g., functions) and validate translations using I/O equivalence, while AlphaTrans~\cite{Ibrahimzada2025AlphaTrans} introduces dependency-aware ordering over fragment-level units. While these methods improve scalability, they continue to treat translation as a collection of independent units, using dependencies primarily as ordering constraints. Interdependent components are still translated in isolation, leading to integration inconsistencies in practice. This limitation is further highlighted by RepoTransBench~\cite{RepoTransBench2024}, which show that strong performance on fine-grained translation does not transfer to repository-level settings, where inter-file dependencies and system-level validation become critical.

In this work, our approach is motivated by a real industrial migration effort involving a large Java-based repository (\textit{STAR}, $\approx$51K LOC). STAR is an internal time-tracking and project management web application (~400 daily users) with a large, extensively tested codebase. The goal of this migration is to adopt Kotlin to benefit from improved type safety (e.g., null-safety), conciseness, and modern language features, while preserving compatibility with the existing JVM-based ecosystem.

In practice, we initially followed a standard file-by-file migration workflow, supported by automated translation tools and iterative compile- and test-driven feedback. However, this approach quickly proved infeasible at scale. Despite repeated refinement, the system frequently failed to compile or pass tests due to inconsistencies across interdependent files. Importantly, these failures were not caused by local translation errors, but by unresolved dependencies and incompatible intermediate states introduced during incremental migration. These observations expose a fundamental limitation of existing workflows: translating files in isolation does not preserve dependency consistency in large, tightly coupled repositories. As a result, migration becomes unstable, requiring significant manual effort to resolve integration issues. This suggests that repository-level migration should be treated as a dependency consistency problem, rather than a collection of independent translation tasks.

We propose \textbf{DepWareTrans}, a dependency aware migration framework that aligns translation units with the repository’s dependency structure. Specifically, our approach groups interdependent files into dependency-consistent batches (i.e., subgraphs of the dependency graph) and performs joint translation of each batch using iterative compile- and test-driven validation. Unlike prior work, we preserve the original test suite and leverage language interoperability to validate correctness throughout the migration process, including in the presence of cyclic dependencies. We evaluate our approach on the industrial repository and multiple open-source systems. Our results show that while file-level baselines achieve reasonable performance on small repositories, they fail to produce a compilable system in large, dependency-rich settings. In contrast, DepWareTrans achieves end-to-end migration with full compilation and test-suite correctness. We further demonstrate that the approach generalizes across interoperable language pairs, including Java--Scala and C\#--F\#.

To guide our study, we address the following research questions:\\
    \noindent\textbf{RQ1:} To what extent do existing file-level and iterative translation workflows satisfy repository-level migration requirements?\\
    \textbf{RQ2:} How effective is dependency-aware batching in improving correctness and scalability?\\
    \textbf{RQ3:} Does the proposed approach generalize across interoperable language pairs?

In summary, this paper makes the following contributions:
\begin{itemize}[leftmargin=*, itemsep=0pt, topsep=2pt, parsep=0pt, partopsep=0pt]
    \item \textbf{Empirical analysis:} We show that file-level migration fails to scale to large, dependency-rich repositories, as most failures arise from dependency inconsistencies rather than translation errors.
    \item \textbf{Methodology:} We propose a dependency-aware batching framework for repository-level migration between interoperable language pairs, enabling joint translation of interdependent components with iterative validation.
    \item \textbf{Practical validation:} We demonstrate the effectiveness of our approach on a large industrial system and multiple open-source repositories.
    \item \textbf{Generalizability:} We show that the approach extends beyond Java--Kotlin to other interoperable language pairs.
\end{itemize}

\section{Related Work}

Recent work has explored the use of LLMs for code translation \cite{CodeFuse13B, Jiao2023Evaluation, Pan2024Lost, Yan2023CodeTransOcean, Yin2024Rectifier}, achieving strong performance on curated benchmarks but exhibiting significant degradation on large, real-world repositories. Repository-scale approaches such as Syzygy \cite{Shetty2024Syzygy} and Oxidizer \cite{Zhang2024Scalable} decompose programs into fine-grained function-level units and validate translations using I/O equivalence. AlphaTrans \cite{Ibrahimzada2025AlphaTrans} translates fragment-level nodes ordered by dependency graphs and relies on translated test suites. Other methods leverage transpilers \cite{Yang2024VERT} or generated specifications \cite{Nitin2024SpecTra} to guide translation, but depend on the availability of language-specific tooling or auxiliary artifacts. Recent benchmarks such as RepoTransBench \cite{RepoTransBench2024} further highlight that repository-level translation remains an open challenge, with low success rates even for state-of-the-art methods. A common limitation across these approaches is the reliance on fragment-level decomposition, where translation units correspond to nodes of the dependency graph and dependencies are handled implicitly via ordering or local validation. This fails to ensure dependency consistency within translation units and often leads to integration errors and ambiguous correctness signals. In contrast, our approach constructs dependency-consistent batches corresponding to subgraphs of the repository dependency graph, enabling coordinated translation of interdependent components and validation against the original test suite for unambiguous correctness.
\vspace{2mm}
\begin{figure*}
    \centering
    \includegraphics[width=0.75\textwidth]{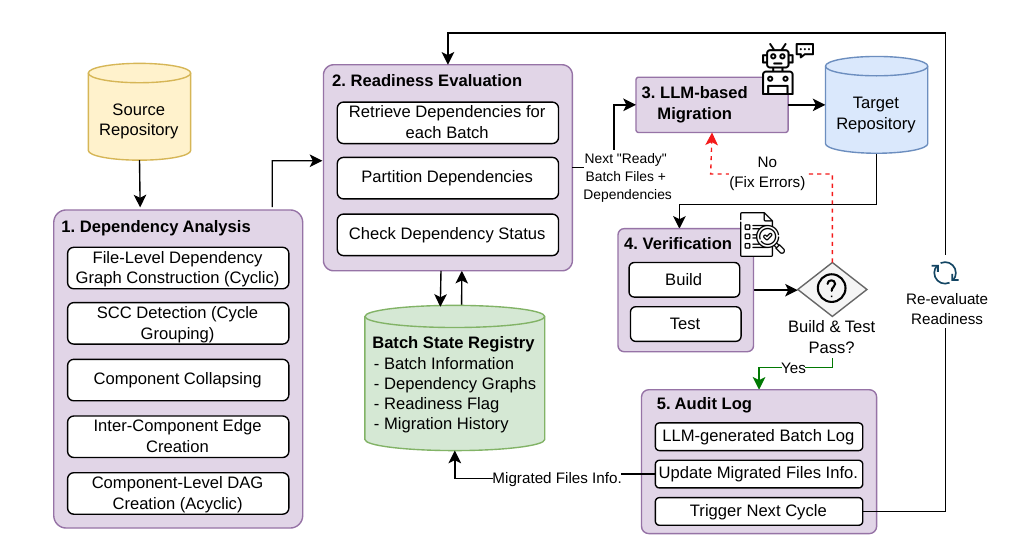}
    \caption{Overview of the dependency-aware migration framework. The process begins with dependency analysis (file-level graph construction and SCC-based cycle handling), followed by readiness evaluation based on dependency status and migration state. Ready batches are translated using an LLM and validated via compilation and testing, with errors iteratively refined through feedback. Sub-steps within each stage are executed in a top-to-bottom order as shown in the figure.}
    \label{fig:metho}
\end{figure*}

Prior work on interoperable language pairs such as Java–Kotlin, Java–Scala, and C\#–F\# has primarily focused on empirical studies of migration and cross-language interaction \cite{KotlinMigrationStudy, KotlinJavaDeps, MigrationExp}, showing that real-world migration is incremental and involves tightly coupled dependencies in mixed-language repositories. However, these works do not propose automated repository-scale migration techniques or address how to structure translation units under such dependency constraints. Our work bridges this gap by explicitly leveraging dependency structure to guide migration and by demonstrating its effectiveness not only on open-source benchmark repositories but also on an industrial codebase.

\section{Methodology}\label{Methodology}
We present the methodology of \textbf{DepWareTrans}, a framework for repository-level migration in interoperable language settings. 
\subsection{Problem Definition}\label{sec:problem}

Let $R$ denote a software repository written in a source programming language $L_s$, consisting of a set of source files $F = \{f_1, f_2, \dots, f_n\}$, where $ f$ denotes an individual file, and $ n$ denotes the total number of files present in the repository. The repository is associated with a build system and a test suite $T$ that defines its functional correctness. Our goal is to migrate $R$ to a target language $L_t$, producing a transformed repository $R'$, while preserving both compilation and behavioral correctness.

We focus on interoperable language pairs that share a common compilation target or runtime (e.g., JVM or .NET), such as Java–Kotlin, Java–Scala, and C\#–F\#. In this setting, components written in $L_s$ and $L_t$ can be compiled and executed jointly, enabling incremental migration with intermediate mixed-language states. This enables \textit{incremental migration}, in which only a subset of files is translated at a time, while the remaining system continues to operate in the original language. A key challenge in this setting arises from the \textit{dependency structure} of the repository. Source files are interconnected through type references, method invocations, inheritance relationships, and module-level interactions. Dependencies can be modeled as a directed graph $G = (F, E)$, where nodes correspond to files and edges represent dependency relationships.

Correctness in repository-scale migration is governed not only by the local correctness of individual translations, but by the \textit{consistency of interdependent components}. Translating files in isolation may lead to unresolved symbols, inconsistent type mappings, or incompatible APIs across files, resulting in compilation failures or incorrect behavior. Therefore, the migration problem can be formulated as identifying an ordering and grouping of files such that translation preserves \textit{dependency consistency}, enabling the system to remain compilable and testable throughout the migration.

In this work, we design a migration approach that (i) preserves compilation correctness at each step, (ii) maintains behavioral correctness with respect to the original test suite $T$, and (iii) minimizes integration errors arising from cross-file dependencies.

\subsection{Overview of the Approach}

To address the challenges outlined in Section~\ref{sec:problem}, we propose a \textit{dependency-aware migration framework} that aligns the unit of translation with the repository’s dependency structure. Instead of translating files independently, our approach groups interdependent files into \textit{dependency-consistent batches} and performs coordinated translation over these units. Figure~\ref{fig:metho} presents the end-to-end workflow of the proposed framework. At a high level, the process consists of the following stages:

\noindent\begin{enumerate}[leftmargin=*, itemsep=0pt, topsep=2pt, parsep=0pt, partopsep=0pt]
    \item \textbf{Dependency Analysis:} Construct a file-level dependency graph and identify strongly connected components (SCCs) to capture cyclic relationships.

    \item \textbf{Batch Construction:} Collapse SCCs to form a component-level directed acyclic graph (DAG), which is partitioned into dependency-consistent batches.

    \item \textbf{Readiness Evaluation:} Determine which batches are ready for migration based on dependency readiness and the current migration state.

    \item \textbf{Batch-Level Translation:} Translate each ready batch using an LLM, incorporating context from its dependencies.

    \item \textbf{Iterative Validation:} Compile and test the translated code using the original test suite; errors are used to iteratively refine the translation.

    \item \textbf{Incremental Migration:} Repeat the process iteratively, updating migrated files and re-evaluating dependencies to enable subsequent batches while maintaining executability.
\end{enumerate}
By combining dependency-aware batching with iterative validation, the framework ensures that translation units reflect the structural dependencies of the system. This enables consistent integration, reduces dependency-induced failures, and supports reliable incremental migration in real-world repositories.

\begin{algorithm}[t]
\caption{Dependency-Aware Incremental Repository Migration}
\label{alg:framework}
\small
\begin{algorithmic}[1]

\REQUIRE Source repository $R$, source language $L_s$, target language $L_t$,
test suite $T$
\ENSURE Migrated repository $R'$

\STATE Extract source files
$F \gets \{f_1, f_2, \dots, f_n\}$ from $R$
\STATE Construct file-level dependency graph $G=(F,E)$
\STATE Compute strongly connected components (SCCs) of $G$
\STATE Collapse SCCs to obtain component-level DAG $G'=(C,E')$
\STATE Construct dependency-consistent batches
$\mathcal{B}=\{B_1,B_2,\dots,B_m\}$ from $G'$
\STATE Initialize migrated file set $M \gets \emptyset$
\STATE Initialize migration state registry $\mathcal{S}$

\WHILE{there exists an unmigrated batch in $\mathcal{B}$}
    \STATE $B \gets
    \textsc{SelectReadyBatch}(\mathcal{B},M,\mathcal{S})$
    
    \STATE $ctx \gets
    \textsc{BuildBatchContext}(B,M,R)$
    
    \STATE $\hat{B} \gets
    \textsc{TranslateBatchWithLLM}(B,ctx,L_s,L_t)$

    \WHILE{\TRUE}
        \STATE Integrate $\hat{B}$ into partially migrated repository $R$
        
        \STATE $buildOK \gets \textsc{Build}(R)$
        
        \IF{$buildOK$}
            \STATE $testOK \gets \textsc{Test}(R,T)$
        \ELSE
            \STATE $testOK \gets \FALSE$
        \ENDIF

        \IF{$buildOK \land testOK$}
            \STATE \textbf{break}
        \ELSE
            \STATE $feedback \gets
            \textsc{CollectErrorsAndFailures}(R)$
            
            \STATE $\hat{B} \gets
            \textsc{RefineBatchWithLLM}(
                \hat{B},feedback,ctx)$
        \ENDIF
    \ENDWHILE

    \STATE $M \gets M \cup \{\text{files in }B\}$
    
    \STATE Update migration state registry $\mathcal{S}$ with migrated files,
    dependency status, and validation outcomes
\ENDWHILE

\STATE $R' \gets R$
\RETURN $R'$

\end{algorithmic}
\end{algorithm}

Algorithm~\ref{alg:framework} summarizes the overall dependency-aware migration process. Starting from a source repository, the algorithm constructs a file-level dependency graph, collapses cyclic dependencies into SCCs, and forms dependency-consistent batches. Migration then proceeds incrementally by selecting ready batches, translating them with an LLM, and iteratively refining them based on build and test feedback until the entire repository is migrated.

\begin{figure}
    \centering
    \includegraphics[width=0.48\textwidth]{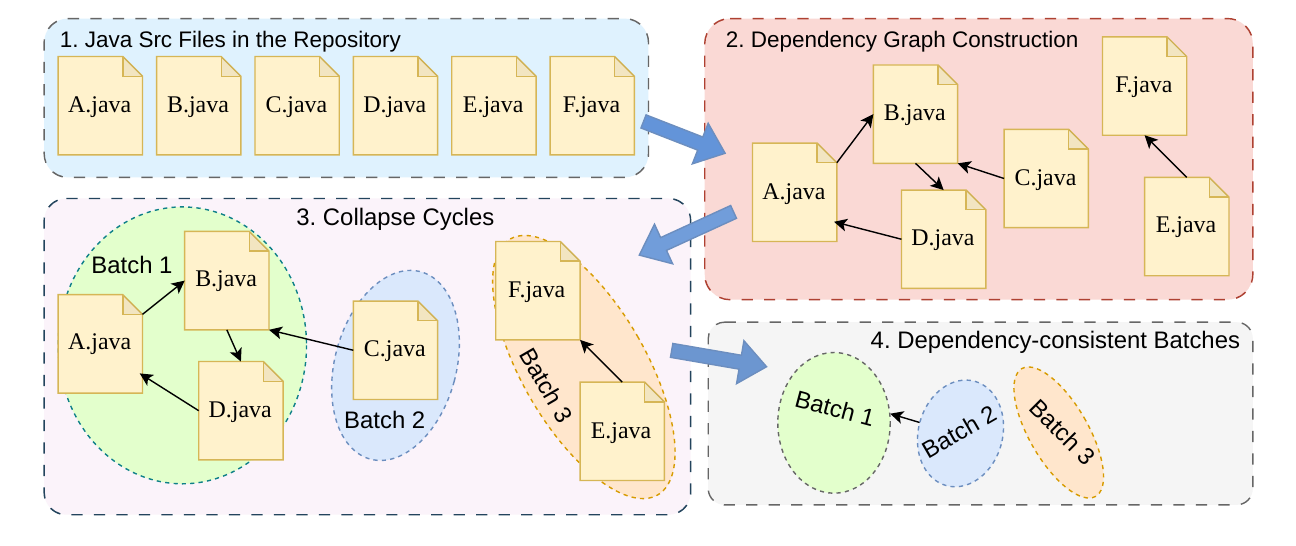}
    \caption{Dependency-aware batch construction. Source files are modeled as a dependency graph; cycles are collapsed into SCCs as atomic units. The graph is then partitioned into dependency-consistent batches, grouping interdependent files for coherent translation.}
    \label{fig:batching}
\end{figure}

\subsection{Dependency Analysis}

Given a repository $R$ with source files $F = \{f_1, f_2, \dots, f_n\}$, we first extract its dependency structure by constructing a directed graph $G = (F, E)$, where each node corresponds to a source file and each edge $(f_i, f_j) \in E$ denotes that file $f_i$ depends on file $f_j$. Dependencies are identified from inter-file relationships such as type references, inheritance, import statements, and method invocations, which also capture dynamic interactions and exceptional control flows across files.

As illustrated in Figure~\ref{fig:batching}, this dependency graph captures the structural coupling between files and reveals both acyclic and cyclic relationships. Cyclic dependencies arise when groups of files are mutually dependent, which is common in real-world systems due to shared abstractions and bidirectional interactions.

To enable structured reasoning, we compute the SCCs\cite{munteanu2018strongly} of $G$. Each SCC represents a maximal set of mutually dependent files and is treated as an atomic unit. We collapse each SCC into a single component node, producing a component-level graph $G' = (C, E')$, where $C$ is the set of components.

By construction, $G'$ is a DAG, as shown in Figure~\ref{fig:metho}, where each node represents a strongly connected component of files. This transformation abstracts cyclic dependencies and provides a component-level dependency structure that serves as the foundation for subsequent batch construction.

\subsection{Dependency-Aware Batch Construction}\label{sec:dependency-aware}
Formally, a \textit{dependency-consistent batch} is a set of files induced by one or more components in the component-level DAG \(G'=(C,E')\), such that: (i) mutually dependent files remain within the same batch, and (ii) every dependency leaving the batch is either already resolved by previously migrated batches. Building on the component-level DAG $G' = (C, E')$ obtained from the dependency analysis phase, we construct \textit{dependency-consistent batches} that serve as the unit of translation. Each batch is constructed based on the component structure of $G'$ and ultimately consists of a set of source files. SCCs are treated as atomic units during batching unless further split due to size constraints and are designed to preserve dependency consistency within the translation unit.

The key objective is to group components such that most dependencies are resolved within the batch, thereby minimizing cross-batch dependencies and reducing integration inconsistencies. As illustrated in Figure~\ref{fig:batching}, this results in batches that align with the structural organization of the repository.

\textbf{Batch Formation.} We leverage the acyclic structure of $G'$ to guide batch construction. Since $G'$ admits a topological ordering, we iteratively group components whose dependencies are either:
(i) already satisfied by previously migrated batches, or  
(ii) contained within the current batch. This grouping strategy approximates \textit{dependency closure}, ensuring that interdependent components are translated together.

\textbf{Handling Cyclic Dependencies.} Since cyclic dependencies are resolved during the dependency analysis phase via SCC collapsing, each component in $G'$ already represents a dependency-consistent unit. As a result, mutually dependent files are guaranteed to be included within the same batch, avoiding inconsistencies that arise from independent translation.

\textbf{Batch Readiness.} At each step, we evaluate the \textit{readiness} of a batch based on its dependencies, as illustrated in Figure~\ref{fig:metho}. A batch is considered ready for migration if all its external dependencies have been:
(a) migrated in previous steps, or (b) included within the batch itself. This readiness criterion enables incremental migration while preserving system correctness.

\textbf{Discussion.} Unlike prior approaches\cite{Ibrahimzada2025AlphaTrans,RepoTransBench2024} that operate on individual files or fine-grained fragments, our method constructs translation units corresponding to subgraphs of the dependency graph. This shift from node-level to subgraph-level translation ensures structural consistency across interdependent components and significantly reduces dependency-induced errors during migration.

\subsection{Batch Translation and Validation}\label{sec:translation}

Once a batch is determined to be ready, we perform \textit{batch-level translation} using a LLM. Unlike file-level approaches, where each file is translated independently, our method translates all files within a batch in a coordinated manner, providing contextual information about related components. The batch-level translation and validation process corresponds to the LLM-based migration and verification stages shown in Figure~\ref{fig:metho}.

\textbf{Batch-Level Translation.} For each batch, we construct an input context that includes: 
(a) the source code of files within the batch, and 
(b) relevant dependency information, such as referenced types, method signatures, and imported interfaces. 
Bounded batch size ensures the input context remains within the model’s input limits.

\textbf{Iterative Validation.} After translating a batch, we validate the result through compilation and test execution, as shown in Figure~\ref{fig:metho}. Specifically, we: (1) attempt to compile the partially migrated repository, and (2) execute the original test suite $T$ to verify behavioral correctness. This is enabled by the interoperability of the source and target languages, which allows translated and untranslated components to be compiled and executed jointly. If compilation or test failures occur, the translation is refined based on the feedback from these errors. This process is repeated iteratively until the batch satisfies compilation and test requirements. Since translation is performed at the batch level, error resolution can account for interactions across multiple files, enabling more effective fixes than isolated file-level refinement.

\subsection{Incremental Migration Workflow}

Our framework performs migration in an incremental manner, translating the repository batch by batch while maintaining system executability throughout the process. At each iteration, a ready batch is selected based on dependency constraints (Section~\ref{sec:dependency-aware}) and translated using the procedure described in Section~\ref{sec:translation}. Once validated, the batch is integrated into the target repository, and the migration state is updated. This includes recording migrated components, updating dependency status, and logging translation outcomes, as illustrated in Figure~\ref{fig:metho}.

\textbf{Interoperability.} A key enabler of incremental migration is the interoperability between the source and target languages. Components written in $L_s$ and $L_t$ can coexist and interact, allowing partially migrated systems to compile and execute correctly. This property ensures that validation can be performed continuously during migration.

\textbf{Migration State Management.} We maintain a migration state that tracks: which components have been translated, the dependency status of remaining components, and validation outcomes for each batch. This state is used to guide batch readiness evaluation and to ensure that dependencies are resolved before translation.

\textbf{Termination.} The process continues until all batches have been translated, resulting in a fully migrated repository $R'$ that satisfies compilation and test correctness criteria.

\textbf{Discussion.} By structuring migration as an incremental process, our approach avoids the risks associated with monolithic translation. Instead, it enables continuous validation and controlled integration, making it suitable for large, real-world repositories.

\section{Experimental Setup}

\textbf{Subjects.} We evaluate our approach on both open-source and industrial repositories. For the Java$\rightarrow$Kotlin migration (RQ1 and RQ2), we consider six widely used Apache Commons projects and a large industrial system (STAR). The open-source repositories range from 6K to 29K LOC and exhibit high test coverage (up to 99\% line coverage). Detailed statistics, including test coverage, are reported in Table~\ref{tab:rq1-baseline-full}. These repositories are widely used in prior software engineering studies as evaluation subjects or benchmarks~\cite{yang2023coderepresentation,Ibrahimzada2025AlphaTrans,hilton}. For RQ3, we extend the evaluation to additional interoperable language pairs, including Java$\rightarrow$Scala and C\#$\rightarrow$F\#. We consider representative repositories such as Commons Text~\cite{commons-text}, GuardClauses~\cite{guardclauses}, MediatR~\cite{mediatr}, and csharp-mcp~\cite{csharp-mcp}, enabling assessment across both ecosystems and models.

\textbf{Industrial System.} STAR is an internal web-based time-tracking and project management system with approximately 400 daily users across multiple organizational roles. The system has been under continuous development since 2018 and supports functionalities such as time entry and approval workflows, leave management, project assignment and effort tracking, and automated billing integration. It includes a role-based access control system and integrates with external accounting systems. The codebase comprises 51K LOC and 1,470 Java files organized into multiple functional modules. Importantly, STAR includes a comprehensive test suite, with test code exceeding production code in size, enabling robust validation of behavioral correctness during migration.

\textbf{Baselines and Variants.} We compare three configurations:

\begin{itemize}[itemsep=0pt, topsep=2pt, parsep=0pt, partopsep=0pt]
    \item \textbf{File-by-file}: Each source file is translated independently without feedback.
    \item \textbf{File-by-file + Feedback}: Each file is translated iteratively, incorporating compiler and test feedback, with a maximum of 10 attempts per file.
    \item \textbf{DepWareTrans (ours)}: Translates dependency-consistent batches with iterative feedback.
\end{itemize}

For ablation, we also evaluate \textbf{DepWareTrans (w/o Feedback)}, which performs dependency-aware batching without iterative refinement.

\textbf{Migration Workflow.} In the file-by-file setting, each Java file is translated independently while the rest of the repository remains unchanged. After each translation attempt, we compile the entire repository; if compilation succeeds, we execute the test suite. Failures trigger a feedback loop: compiler and test errors are provided to the LLM for refinement. Each file is retried up to 10 iterations.

In our approach, we first construct a dependency graph using static analysis based on imports, type references, and package structure. We then group mutually dependent files into batches (maximum 25 files per batch) using strongly connected components and process batches in a dependency-consistent order. Each batch is translated jointly, compiled, and tested. If failures occur, we iteratively refine the batch using feedback until success.

\textbf{Metrics.} We evaluate translation effectiveness using:

\begin{itemize}[itemsep=0pt, topsep=2pt, parsep=0pt, partopsep=0pt]
    \item \textbf{Compile Success (\%)}: Percentage of units (files or batches) that successfully compile.
    \item \textbf{Test Success (\%)}: Percentage of units that both compile and pass all tests.
\end{itemize}

Because the migration unit differs across methods (files versus dependency-consistent batches), all methods are compared using the same external criterion: after each migration step, we rebuild the entire partially migrated repository and run the full test suite. Thus, correctness is assessed at the repository state level rather than by local unit validity.

\textbf{Implementation Details.} All Java-based experiments are conducted using Maven (version 3.9.11) with Java 17. Kotlin translation is performed using Kotlin 1.9.25 via Maven plugins. For .NET experiments, we use .NET SDK 10.0.103 with runtime support for .NET 8 and 10. Scala experiments are executed within the Maven toolchain.

We use GPT-5.3 Codex (medium reasoning setting, default parameters) for all experiments. To assess generalizability across models, we additionally evaluate the Java$\rightarrow$Scala setting using Claude Opus 4.6 and GPT-5.4. No hyperparameter tuning is performed.

The prompting strategy differs by configuration. In the file-by-file baseline, the model receives a single source file and produces a translated version. In our approach, the model operates on dependency-consistent batches, where the full source code of all files in the batch is included in the prompt, together with selected dependency context (e.g., referenced types, method signatures, and interfaces) from related files, rather than the entire dependency closure. In feedback-based settings, compiler and test errors from previous attempts are incorporated into the prompt to guide refinement. Full prompting templates and scripts are available in our artifact repository. 

\textbf{Stopping Criteria.} In the baseline, each file is retried up to 10 iterations to avoid unbounded execution. In our approach, no fixed iteration limit is imposed; however, we observe that convergence typically occurs within a small number of iterations, and only one case exceeds 10 iterations as shown in Table \ref{tab:star-iteration-success}.
\vspace{2mm}
\begin{table*}[t]
\centering
\caption{Performance of file-level migration approaches across repositories.}
\label{tab:rq1-baseline-full}
\begin{tabular}{l r r rr rr rr}
\toprule
& \multicolumn{2}{c}{\textbf{Size}} & \multicolumn{2}{c}{\textbf{Test Coverage}} & \multicolumn{2}{c}{\textbf{File-by-file}} & \multicolumn{2}{c}{\textbf{File-by-file + Feedback}} \\
\cmidrule(lr){2-3} \cmidrule(lr){4-5} \cmidrule(lr){6-7} \cmidrule(lr){8-9}
\textbf{Repository / Feature} & LOC & \#Java & Line (\%) & Branch (\%) & Compile (\%) & Test (\%) & Compile (\%) & Test (\%) \\
\midrule
\multicolumn{9}{l}{\textit{Open-source repositories}} \\
\midrule
commons-cli\cite{commons-cli} & 9,718 & 36 & 95.90 & 94.86 & 69.44 & 63.89 & 100.00 & 100.00 \\
commons-codec\cite{commons-codec} & 25,431 & 87 & 94.89 & 90.23 & 80.46 & 68.97 & 100.00 & 100.00 \\
commons-csv\cite{commons-csv} & 6,295 & 12 & 99.59 & 97.59 & 50.00 & 41.67 & 58.33 & 58.33 \\
commons-exec\cite{commons-exec} & 5,177 & 37 & 74.75 & 57.78 & 86.49 & 81.08 & 100.00 & 97.30 \\
commons-fileupload\cite{commons-fileupload} & 7,094 & 45 & 57.20 & 52.34 & 80.00 & 80.00 & 97.78 & 97.78 \\
commons-text\cite{commons-text} & 29,742 & 112 & 99.38 & 97.82 & 29.46 & 8.04 & 29.46 & 8.04 \\
\midrule
\multicolumn{9}{l}{\textit{Industrial repository (STAR) by functional feature}} \\
\midrule
urlaub & 7,896 & 213 & -- & -- & 19.72 & 4.23 & 22.07 & 4.69 \\
stammdaten & 7,023 & 211 & -- & -- & 14.69 & 1.42 & 17.06 & 1.42 \\
leistungsnachweis & 5,608 & 199 & -- & -- & 24.62 & 4.02 & 24.62 & 4.02 \\
auftrag & 5,400 & 180 & -- & -- & 20.00 & 1.11 & 54.44 & 1.11 \\
wochenarbeitszeit & 4,423 & 128 & -- & -- & 25.00 & 11.72 & 96.88 & 16.41 \\
krankheit & 4,844 & 114 & -- & -- & 11.40 & 0.88 & 11.40 & 0.88 \\
common & 2,283 & 74 & -- & -- & 9.46 & 8.11 & 10.81 & 8.11 \\
persistence & 2,400 & 69 & -- & -- & 55.07 & 55.07 & 59.42 & 59.42 \\
zeiterfassung & 2,587 & 68 & -- & -- & 17.65 & 0.00 & 100.00 & 0.00 \\
stundenuebersicht & 1,716 & 59 & -- & -- & 15.25 & 1.69 & 15.25 & 1.69 \\
mitarbeiterimport & 768 & 27 & -- & -- & 11.11 & 3.70 & 11.11 & 3.70 \\
populators & 1,601 & 27 & -- & -- & 48.15 & 48.15 & 48.15 & 48.15 \\
datapopulator & 1,693 & 26 & -- & -- & 65.38 & 57.69 & 69.23 & 61.54 \\
config & 1,196 & 21 & -- & -- & 57.14 & 38.10 & 57.14 & 38.10 \\
mitarbeiter & 501 & 15 & -- & -- & 13.33 & 0.00 & 13.33 & 0.00 \\
abwesenheitsuebersicht & 263 & 12 & -- & -- & 8.33 & 0.00 & 100.00 & 16.67 \\
nonprod-root & 406 & 11 & -- & -- & 18.18 & 18.18 & 18.18 & 18.18 \\
dbcleanup & 219 & 10 & -- & -- & 20.00 & 0.00 & 20.00 & 0.00 \\
cloudwatch & 180 & 5 & -- & -- & 60.00 & 40.00 & 60.00 & 40.00 \\
backend-root & 13 & 1 & -- & -- & 100.00 & 100.00 & 100.00 & 100.00 \\
\midrule
STAR (overall) & 51,020 & 1,470 & 96.47 & 93.44 & 22.11 & 8.50 & 38.16 & 9.39 \\
\bottomrule
\end{tabular}
\end{table*}

\section{Results}\label{sec:results}
In this section, we discuss the study results by posing and answering three research questions.

\subsection{RQ1: Limitations of File-Level Migration}

To investigate how file-level migration strategies perform in real-world scenarios, we evaluate their effectiveness, with and without iterative feedback, across a diverse set of open-source repositories and an industrial system (STAR). The results are summarized in Table~\ref{tab:rq1-baseline-full}. We report coverage at the repository level, as tests are defined and executed globally rather than per feature. Since correctness is evaluated based on full-system compilation and test execution, repository-level coverage is the relevant metric.

\begin{table}[t]
\centering
\caption{Shared diagnostic signatures across failure-prone repositories (post-feedback).}
\label{tab:rq1-error-alignment}
\resizebox{\columnwidth}{!}{%
\begin{tabular}{l rr rr}
\toprule
& \multicolumn{2}{c}{STAR} & \multicolumn{2}{c}{commons-text} \\
\cmidrule(lr){2-3} \cmidrule(lr){4-5}
Error Signature & Count & Share (\%) & Count & Share (\%) \\
\midrule
\texttt{cannot find symbol}~\cite{oracle-javac} & 12,994 & 96.98 & 720 & 69.90 \\
\texttt{package ... does not exist}~\cite{oracle-javac} & 357 & 2.66 & 70 & 6.80 \\
\texttt{ClassNotFoundException}~\cite{oracle-classnotfound} & 0 & 0.00 & 240 & 23.30 \\
\texttt{incompatible types}~\cite{oracle-javac} & 27 & 0.20 & 0 & 0.00 \\
\texttt{method ... cannot be applied}~\cite{oracle-javac} & 1 & 0.01 & 0 & 0.00 \\
Generic test failure (\texttt{Tests run ...})~\cite{maven-surefire-errors} & 10 & 0.07 & 0 & 0.00 \\
\bottomrule
\end{tabular}
}
\end{table}

On smaller open-source repositories, file-by-file translation achi-eves moderate performance, with compilation success ranging from 50.00\% to 86.49\% and test success from 41.67\% to 81.08\%. Incorporating iterative feedback significantly improves results in several cases, achieving 100\% compilation and test success for projects such as \textit{commons-cli} and \textit{commons-codec}. However, this improvement is not consistent across all repositories. For instance, on \textit{commons-csv}, performance remains limited (58.33\% compile and test success), and on \textit{commons-text}, both approaches fail substantially, with only 29.46\% compilation and 8.04\% test success even after feedback. 

The limitations of file-level approaches become more pronounced in the industrial repository. Across its functional modules, file-by-file translation consistently yields low success rates, with compilation often below 25\% and test success frequently near zero. While feedback improves compilation in some cases (e.g., from 25.00\% to 96.88\% for \textit{wochenarbeitszeit}), it fails to translate into corresponding test success, which remains low (16.41\%). Overall, at the repository level, performance improves only marginally from 22.11\% to 38.16\% in compilation and from 8.50\% to 9.39\% in test success, indicating that feedback alone is insufficient for large, interdependent systems.

 To better understand these failures, we analyze the distribution of error types, as shown in Table~\ref{tab:rq1-error-alignment}. We observed that failures in the two most problematic repositories are not random translation mistakes, but are overwhelmingly dependency-linkage failures. In \textit{STAR}, nearly all failed attempts correspond to compile-time reference breakages (e.g., missing symbols and packages), with only a negligible fraction attributable to type or API incompatibilities. In \textit{commons-text}, a similar structural pattern emerges: compile-time linkage failures dominate, while the remaining errors are largely runtime linkage issues (e.g., \texttt{ClassNotFoundException}), which still reflect failures at dependency boundaries rather than local translation quality.

This cross-repository alignment makes the underlying failure mode explicit. When grouped by error family, both repositories concentrate in the same dependency-related classes, indicating a shared structural limitation. Under file-by-file migration, translating a file while its dependency neighborhood remains unmigrated creates unstable mixed-language boundaries, leading to cascading unresolved references and classpath inconsistencies.
\vspace{2mm}
\begin{table*}[t]
\centering
\caption{Effectiveness of dependency-aware batching and feedback on repositories where file-level approaches underperform.}
\label{tab:rq2-merged}
\begin{tabular}{l rr rr rrr}
\toprule
& \multicolumn{2}{c}{File-by-file + Feedback} 
& \multicolumn{2}{c}{DepWareTrans (w/o Feedback)} 
& \multicolumn{3}{c}{DepWareTrans} \\
\cmidrule(lr){2-3} \cmidrule(lr){4-5} \cmidrule(lr){6-8}
Repository 
& Compile (\%) & Test (\%)
& Compile (\%) & Test (\%)
& Compile (\%) & Test (\%) & Avg. Iter. \\
\midrule

Commons-csv\cite{commons-csv} & 58.33 & 58.33 & 66.66 & 66.66 & 91.67 & 91.67 & 0.80 \\

Commons-text\cite{commons-text} & 29.46 & 8.04 & 100.00 & 100.00 & 100.00 & 100.00 & 0.00 \\

STAR & 38.16 & 9.39 & 7.84 & 7.84 & 100.00 & 100.00 & 3.45 \\

\bottomrule
\end{tabular}
\end{table*}

\begin{table*}[t]
\centering
\caption{STAR cumulative success rate by feedback iteration budget $k$.}
\label{tab:star-iteration-success}

\begin{tabular}{l rrrrrrrrrr}
\toprule
$k$ & 0 & 1 & 2 & 3 & 4 & 5 & 6 & 7 & 8 & 20 \\
\midrule
Compile Success (\%) 
& 7.84 & 33.33 & 62.75 & 88.24 & 96.08 & 100.00 & 100.00 & 100.00 & 100.00 & 100.00 \\
Compile+Test Success (\%) 
& 7.84 & 15.69 & 31.37 & 49.02 & 66.67 & 82.35 & 84.31 & 96.08 & 98.04 & 100.00 \\
\bottomrule
\end{tabular}

\end{table*}

These observations suggest that the primary limitation of baseline approaches is not the translation capability of the model, but the choice of migration unit, which ignores the dependency structure of the repository. This motivates a shift from file-level translation to dependency-aware units (e.g., SCC- or component-level batches), which co-migrate interdependent files and directly target the dominant source of failure identified in RQ1.

\answerbox{%
\textbf{Answer to RQ1:} File-level migration does not scale to dependency-rich repositories; even with iterative feedback, cross-file dependency inconsistencies remain unresolved, resulting in low end-to-end correctness.
}

\subsection{RQ2: Effectiveness of DepWareTrans}

Motivated by the limitations identified in RQ1, where file-level approaches fail in dependency-rich repositories, we evaluate the effectiveness of our proposed dependency-aware migration framework. Specifically, we compare three configurations: (i) file-by-file migration with feedback, (ii) dependency-aware batching without feedback, and (iii) the full approach combining batching with iterative feedback. The results are summarized in Table~\ref{tab:rq2-merged}, with convergence behavior further analyzed in Table~\ref{tab:star-iteration-success}. 

We focus on repositories where file-by-file approaches underperform, as these cases expose the limitations of file-level migration. Repositories where file-level methods already achieve near-perfect results (e.g., commons-cli, commons-codec) provide limited insight into failure modes and are therefore not the focus of this analysis.

Across all repositories, we observe that dependency-aware batching significantly improves migration outcomes compared to file-level workflows. For example, \textit{commons-text}, batching alone is sufficient to fully resolve migration failures, achieving 100\% compilation and test success without any feedback, compared to only 29.46\% and 8.04\% under file-by-file migration with feedback. Consistent with the findings in RQ1, this indicates that failures in this repository are predominantly structural, arising from inconsistent handling of interdependent components. Once dependency consistency is enforced through batching, no further refinement is required. In contrast, \textit{commons-csv} exhibits a mixed failure mode. Batching without feedback improves performance over the file-level baseline (66.66\% vs. 58.33\%), and the full approach further increases success to 91.67\%. However, unlike \textit{commons-text}, batching combined with feedback does not achieve complete success. Manual inspection shows that the remaining failure is concentrated in a single large, central class, approximately 3K LOC, with complex builder logic and strict API/ABI expectations used widely across the repository. This case suggests that, while dependency-aware batching resolves most repository-level inconsistencies, some architectural hub classes require more than iterative LLM-based repair. In particular, large, highly coupled classes may need complementary decomposition, refactoring, or manual guidance before they can be migrated reliably.

For the industrial repository \textit{STAR}, batching alone is insufficient, achieving only 7.84\% success, equivalent to the zero-iteration baseline. This does not contradict the RQ1 finding that STAR failures are primarily dependency-related: the file-level baseline in RQ1 also used iterative feedback, yet still achieved only 38.16\% compilation and 9.39\% test success. Rather, the result shows that feedback is ineffective when applied to isolated files under inconsistent dependency boundaries. Dependency-aware batching first establishes a structurally consistent migration unit; once this consistency is in place, feedback can effectively resolve the remaining translation and integration errors that arise at scale.

When combined with iterative feedback, the full approach achie-\\ves 100\% compilation and test success, substantially outperforming the file-level baseline. Table~\ref{tab:star-iteration-success} further illustrates this convergence behavior. Starting from a low baseline of 7.84\% at $k=0$, performance improves steadily with additional feedback iterations, reaching 49.02\% at $k=3$, 82.35\% at $k=5$, and 98.04\% by $k=8$. Notably, most improvements occur within the first few iterations, suggesting that once dependency consistency is established, remaining errors become localized and can be corrected efficiently through feedback.

Overall, these results demonstrate that batching and feedback address complementary aspects of the migration problem. Batching resolves structural inconsistencies caused by interdependent components, while iterative feedback corrects residual translation and integration errors. However, the results on \textit{commons-csv} also indicate that extremely large or highly central components may require additional handling beyond the current approach. 

\answerbox{%
\textbf{Answer to RQ2:} Dependency-aware batching preserves dependency coherence, while iterative feedback fixes residual errors; together, they enable reliable end-to-end migration.
}

\subsection{RQ3: Generalization Across Language Pairs}
\vspace{2mm}
\begin{table*}[t]
\centering
\caption{Generalization across interoperable language pairs, comparing file-by-file + feedback with DepWareTrans variants.}
\label{tab:rq3-folder-summary}
\resizebox{2\columnwidth}{!}{%
\begin{tabular}{l l l rr rr rrr}
\toprule
& & & \multicolumn{2}{c}{\makecell{File-by-file \\ + Feedback}}
& \multicolumn{2}{c}{\makecell{DepWareTrans \\ (w/o Feedback)}} 
& \multicolumn{3}{c}{DepWareTrans} \\
\cmidrule(lr){4-5} \cmidrule(lr){6-7} \cmidrule(lr){8-10}
Repository & \makecell{Source$\rightarrow$ \\ Target} & LLM Used
& Compile (\%) & Test (\%)
& Compile (\%) & Test (\%)
& Compile (\%) & Test (\%) & Avg. Iter. \\
\midrule

GuardClauses\cite{guardclauses} & C\#$\rightarrow$F\# & Codex 5.3
& 100.00 & 100.00
& 40.00 & 40.00
& 100.00 & 100.00 & 2.60 \\

MediatR\cite{mediatr} & C\#$\rightarrow$F\# & Codex 5.3
& 41.86 & 41.86
& 100.00 & 66.67
& 100.00 & 100.00 & 0.00 \\

csharp-mcp\cite{csharp-mcp} & C\#$\rightarrow$F\# & Codex 5.3
& 66.67 & 33.33
& 0.00 & 0.00
& 100.00 & 100.00 & 1.00 \\

\midrule

Commons Text\cite{commons-text} & Java$\rightarrow$Scala & Codex 5.3
& 29.46 & 8.04
& 78.56 & 59.82
& 100.00 & 100.00 & 0.68 \\

Commons Text\cite{commons-text} & Java$\rightarrow$Scala & Claude Opus 4.6
& 26.78 & 8.04
& 47.37 & 31.58
& 100.00 & 100.00 & 0.89 \\

Commons Text\cite{commons-text} & Java$\rightarrow$Scala & GPT-5.4
& 24.10 & 7.14
& 5.88 & 5.88
& 100.00 & 100.00 & 5.59 \\

\bottomrule
\end{tabular}
}
\end{table*}

We evaluate whether the effectiveness of our approach generalizes beyond Java--Kotlin to other interoperable language ecosystems and across different underlying LLMs. Table~\ref{tab:rq3-folder-summary} summarizes the results across multiple repositories, language pairs, and models.

We first examine the behavior of file-by-file migration with feedback across language pairs. The results show substantial variability in performance depending on the repository and language ecosystem. For example, file-level migration achieves 100\% success in \textit{GuardClauses} (C\#→F\#), but drops to 66.67\% compile and 33.33\% test success in \textit{csharp-mcp}, and further to 29.46\% compile and 8.04\% test success in \textit{Commons Text} (Java→Scala). This variability indicates that file-by-file workflows are highly sensitive to repository structure and dependency complexity, and do not generalize reliably across language pairs.

Dependency-aware batching without feedback provides more consistent improvements, but its effectiveness remains limited and repository-dependent. For instance, batching alone achieves 40.00\% success in \textit{GuardClauses}, 47.37\%/31.58\% in \textit{Commons Text}, and fails entirely in \textit{csharp-mcp} (0.00\%). These results are consistent with RQ2, suggesting that while batching resolves structural inconsistencies, it does not address residual translation errors or complex semantic interactions. In contrast, the full approach combining batching with iterative feedback consistently achieves 100\% compilation and test success across all repositories, language pairs, and models. This includes both JVM-based (Java→Scala) and .NET-based (C\#→F\#) ecosystems, as well as different LLMs (Codex 5.3, GPT-5.4 and Claude Opus 4.6). Notably, the approach achieves these results even when file-level workflows fail significantly, as in \textit{Commons Text} and \textit{csharp-mcp}. These results highlight two key aspects of generalization. First, the limitations of file-by-file migration are consistent across language pairs: performance degrades in repositories with complex dependency structures, regardless of the specific languages involved. Second, the improvements introduced by DepWareTrans are stable across both ecosystems and models, indicating that the gains arise from the migration strategy, specifically, enforcing dependency consistency and iterative refinement rather than model-specific capabilities.

Overall, these findings suggest that dependency-aware batching combined with feedback provides a robust and generalizable solution for repository-level migration in interoperable language settings, while file-level approaches remain brittle and highly sensitive to repository structure.

\answerbox{%
\textbf{Answer to RQ3:} Across the evaluated interoperable language pairs and LLMs, DepWareTrans generalizes effectively: file-by-file migration shows unstable performance, whereas dependency-aware batching with iterative feedback consistently achieves full correctness in our experiments.
}

\section{Discussion}

 \textbf{Why Dependency-Aware Migration Works.} The results across RQ1–RQ3 indicate that dependency inconsistency is the dominant bottleneck in repository-level code translation. 

\textit{Problem.} File-level translation violates implicit invariants across interdependent components. In large repositories, classes are tightly coupled through shared data models, inheritance, and APIs. Translating such components independently produces inconsistent intermediate states, leading to unresolved references, type mismatches, and linkage errors. As a result, iterative feedback alone is insufficient: it can correct local errors, but cannot restore consistency across independently translated components.

\textit{Solution.} Dependency-aware batching addresses this by enforcing \emph{local consistency closure}. By co-migrating interdependent files, it ensures that symbols, types, and interfaces remain consistent within the translation unit. This explains the behavior observed in \textit{commons-text}, where failures are primarily structural and are fully resolved by batching alone. However, structural consistency is not sufficient in all cases. In the industrial repository \textit{STAR}, batching must be combined with iterative feedback to achieve full correctness. This indicates that, once dependency inconsistencies are removed, remaining errors stem from translation-level issues such as control flow, API usage, and language-specific constructs. Feedback complements batching by refining these aspects. The results for \textit{commons-csv} highlight a limitation of the approach. While batching and feedback improve success rates, they struggle with large, central components that act as architectural hubs. In such cases, translation errors propagate widely due to strong coupling and strict API constraints. Addressing these scenarios may require additional techniques, such as specialized handling of large classes or interface-driven translation.

Overall, repository-level migration can be viewed as preserving cross-component invariants under dependency constraints. Dependency-aware batching addresses the structural dimension of this problem, while iterative feedback resolves residual errors, together enabling reliable migration in practice.

\textbf{Resource and Cost Analysis.} Across the three repositories shared by the file-by-file baseline and DepWareTrans (Table. \ref{tab:rq2-merged}), the clearest resource difference appears in total token consumption rather than overall active time. The file-by-file baseline consumed 16,941,405 total tokens, whereas DepWareTrans consumed 4,990,616, corresponding to a 70.5\% reduction overall. This reduction is consistent across all three repositories: commons-csv drops from 1,047,288 to 97,848 total tokens, commons-text from 5,442,491 to 555,712, and STAR from 10,451,626 to 4,337,056. In contrast, total active time across the same repositories is similar overall, with file-by-file requiring about 21h 59m 32s and DepWareTrans 21h 52m 16s; however, the per-repository pattern differs, as DepWareTrans is slower on the smaller OSS repositories (commons-csv: 7m 59s vs. 3m 19s; commons-text: 1h 42m 21s vs. 34m 21s) but modestly faster on STAR (20h 01m 56s vs. 21h 21m 51s). Overall, these results suggest that DepWareTrans’s primary efficiency advantage lies in substantially reducing token consumption, while its time advantage becomes more visible on larger, dependency-heavy systems.

\textbf{Implications for Practitioners.} Our results have direct implications for practitioners performing large-scale language migration. 

File-by-file migration, even when supported by modern LLMs and feedback loops, is insufficient for systems with non-trivial dependency structure. As demonstrated in the industrial case (Table~\ref{tab:rq1-baseline-full}), such approaches yield low compilation and test success rates despite multiple refinement attempts.

Dependency-aware batching provides a practical strategy for incremental migration. By ensuring that interdependent files are migrated together, the approach avoids unstable mixed-language boundaries that lead to cascading failures. Combined with iterative feedback, this enables reliable end-to-end migration while preserving the ability to validate intermediate states through compilation and testing. The iteration results (Table~\ref{tab:star-iteration-success}) suggest that the cost of refinement is manageable in practice. Once dependency consistency is established, most errors are resolved in a few iterations, enabling feasible real-world deployment.

\textbf{Lessons from the Industrial Case.} Our experience with the STAR repository provides several practical insights. First, feedback alone is insufficient at the file level, as it cannot resolve failures caused by inconsistent dependency boundaries. This is evident from the limited improvement between file-by-file and file-by-file with feedback in RQ1.
Dependency-aware batching significantly reduces failure modes, but does not completely eliminate the need for refinement in complex systems. While batching alone achieves full success in simpler repositories, the industrial setting requires iterative feedback to resolve residual translation and integration issues.
Once dependency consistency is enforced, convergence is rapid. As shown in Table~\ref{tab:star-iteration-success}, success rates increase sharply within the first few iterations, indicating that most remaining errors are localized and can be corrected efficiently.

\textbf{Applicability and Scope.} The effectiveness of our approach depends on several key conditions. First, it is only well-suited to interoperable or co-executable language pairs (e.g., JVM- or .NET-based ecosystems), where mixed-language execution allows incremental migration. Second, the approach assumes the availability of a buildable system and an executable test suite, which are necessary for feedback-driven refinement. The approach may be less effective in settings where dependencies cannot be reliably extracted, or where test coverage is insufficient to validate correctness. Additionally, for non-interoperable language pairs, where incremental execution is not feasible, alternative migration strategies may be required.

\textbf{Implications for Research.} Our findings suggest that improving repository-level code translation requires reconsidering both evaluation and methodology. Current approaches that focus on file- or function-level benchmarks do not capture the structural challenges observed in real-world systems. As shown in this work, dependency-aware migration provides a more appropriate abstraction for large-scale translation tasks. Furthermore, the results indicate that structural factors play a critical role in translation success, independent of the underlying LLM. This suggests that future work should focus not only on improving model capabilities but also on designing translation frameworks that explicitly account for dependency structure and system-level consistency.

\section{Threats to Validity}

\textbf{Internal Validity.} This concerns whether observed improvements are attributable to our approach. We mitigate LLM non-determinism by using fixed prompts and default parameters across all experiments, without task-specific tuning, and apply the feedback loop consistently across configurations to ensure fair comparison. Our approach also relies on static dependency extraction (imports, type references, package structure), which may introduce inaccuracies. However, consistent results across repositories suggest robustness to minor inference errors.

\textbf{Construct Validity.} We measure effectiveness using compilation and test success rates, where a unit is considered successful only if it both compiles and passes all tests. While these metrics reflect practical correctness, they depend on the quality and completeness of the test suites; limited coverage may not fully guarantee semantic equivalence. 

\textbf{External Validity.} This concerns the generalizability of our findings. We evaluate our approach on multiple open-source repositories and a large industrial system, across several interoperable language pairs (Java--Kotlin, Java--Scala, and C\#--F\#). However, results may not generalize to non-interoperable language pairs, where incremental migration and mixed-language execution are not feasible. Moreover, the effectiveness of our approach depends on the availability of buildable systems and executable test suites, which may not exist in all real-world repositories. The industrial case study focuses on a single organization and domain, which may limit its generalizability to other contexts.

\textbf{Conclusion Validity.} The consistently large performance differences between baseline and proposed approaches reduce the likelihood that improvements are due to random variation. However, the limited number of repositories evaluated in certain settings (e.g., RQ2 and RQ3) may affect the generality of the conclusions, and further studies would strengthen confidence. 

\section{Conclusion}

This paper revisits repository-level migration as a problem of structural consistency in interoperable language settings. While recent advances in LLMs have improved local code translation, our results show that these gains do not translate to end-to-end correctness when migrating real-world systems incrementally. Across both open-source and industrial repositories, failures primarily arise from inconsistencies at dependency boundaries rather than errors in individual file translations.

We propose \textbf{DepWareTrans}, a dependency-aware migration framework that elevates the unit of translation from isolated files to dependency-consistent batches, combined with iterative compile- and test-driven validation. Our evaluation demonstrates that this approach enables reliable, end-to-end migration in settings where file-level workflows fail, and that the benefits generalize across multiple interoperable language pairs.

These findings suggest that effective repository-level migration in co-executable settings must explicitly incorporate dependency structure into both migration units and validation. This perspective opens new directions for structure-aware transformation, including improved handling of architectural hotspots and tighter integration between dependency analysis and LLM-based translation.\\

\section*{Data Availability Statement}
We release a replication package with scripts, configurations, and open-source repositories, available at \url{https://doi.org/10.6084/m9.figshare.32133676}. While we provide STAR’s migration scripts and results, confidentiality constraints prevent releasing its source code. Access to the STAR codebase may be granted upon reasonable request, subject to organizational approval.

\balance
\bibliographystyle{ACM-Reference-Format}
\bibliography{sample-base}

@article{Ibrahimzada2025AlphaTrans,
  author = {Ibrahimzada, Ali Reza and Ke, Kaiyao and Pawagi, Mrigank and Abid, Muhammad Salman and Pan, Rangeet and Sinha, Saurabh and Jabbarvand, Reyhaneh},
  title = {AlphaTrans: A Neuro-Symbolic Compositional Approach for Repository-Level Code Translation and Validation},
  journal = {Proceedings of the ACM on Software Engineering},
  volume = {2},
  number = {FSE},
  year = {2025},
  publisher = {Association for Computing Machinery},
  address = {New York, NY, USA},
  doi = {10.1145/3729379},
  url = {https://doi.org/10.1145/3729379}
}

@inproceedings{CodeFuse13B,
  author = {Di, Peng and Li, Jianguo and Yu, Hang and Jiang, Wei and Cai, Wenting and Cao, Yang and Chen, Chaoyu and Chen, Dajun and Chen, Hongwei and Chen, Liang and Fan, Gang and Gong, Jie and Gong, Zi and Hu, Wen and Guo, Tingting and Lei, Zhichao and Li, Ting and Li, Zheng and Liang, Ming and Liao, Cong and Liu, Bingchang and Liu, Jiachen and Liu, Zhiwei and Lu, Shaojun and Shen, Min and Wang, Guangpei and Wang, Huan and Wang, Zhi and Xu, Zhaogui and Yang, Jiawei and Ye, Qing and Zhang, Gehao and Zhang, Yu and Zhao, Zelin and Zheng, Xunjin and Zhou, Hailian and Zhu, Lifu and Zhu, Xianying},
  title = {CodeFuse-13B: A Pretrained Multi-lingual Code Large Language Model},
  booktitle = {Proceedings of the 46th International Conference on Software Engineering: Software Engineering in Practice (ICSE-SEIP)},
  year = {2024},
  pages = {418--429},
  publisher = {ACM},
  address = {New York, NY, USA},
  doi = {10.1145/3639477.3639719}
}

@inproceedings{Jiao2023Evaluation,
  author = {Jiao, Mingsheng and Yu, Tingrui and Li, Xuan and Qiu, Guanjie and Gu, Xiaodong and Shen, Beijun},
  title = {On the Evaluation of Neural Code Translation: Taxonomy and Benchmark},
  booktitle = {Proceedings of the 38th IEEE/ACM International Conference on Automated Software Engineering (ASE)},
  year = {2023},
  pages = {1529--1541},
  isbn = {9798350329964},
  publisher = {IEEE Press},
  url = {https://doi.org/10.1109/ASE56229.2023.00114},
  doi = {10.1109/ASE56229.2023.00114}
}

@inproceedings{Yan2023CodeTransOcean,
  author = {Yan, Weixiang and Tian, Yuchen and Li, Yunzhe and Chen, Qian and Wang, Wen},
  title = {CodeTransOcean: A Comprehensive Multilingual Benchmark for Code Translation},
  booktitle = {Findings of the Association for Computational Linguistics: EMNLP},
  year = {2023},
  pages = {5067--5089},
  doi = {10.18653/v1/2023.findings-emnlp.337}
}

@article{Yin2024Rectifier,
  title={Rectifier: Code translation with corrector via llms},
  author={Yin, Xin and Ni, Chao and Nguyen, Tien N and Wang, Shaohua and Yang, Xiaohu},
  journal={arXiv preprint arXiv:2407.07472},
  year={2024},
  doi = {
https://doi.org/10.48550/arXiv.2407.07472}
}

@article{shetty2024syzygy,
  title={Syzygy: Dual code-test c to (safe) rust translation using llms and dynamic analysis},
  author={Shetty, Manish and Jain, Naman and Godbole, Adwait and Seshia, Sanjit A and Sen, Koushik},
  journal={arXiv preprint arXiv:2412.14234},
  year={2024},
  doi ={
https://doi.org/10.48550/arXiv.2412.14234}
}

@article{Zhang2024Scalable,
author = {Zhang, Hanliang and David, Cristina and Wang, Meng and Paulsen, Brandon and Kroening, Daniel},
title = {Scalable, Validated Code Translation of Entire Projects using Large Language Models},
year = {2025},
issue_date = {June 2025},
publisher = {Association for Computing Machinery},
address = {New York, NY, USA},
volume = {9},
number = {PLDI},
url = {https://doi.org/10.1145/3729315},
doi = {10.1145/3729315},
journal = {Proc. ACM Program. Lang.},
month = jun,
articleno = {212},
numpages = {26}
}

@article{Yang2024VERT,
  author = {Yang, Aidan ZH and Takashima, Yoshiki and Paulsen, Brandon and Dodds, Josiah and Kroening, Daniel},
  title = {VERT: Verified Equivalent Rust Transpilation with Large Language Models as Few-Shot Learners},
  journal = {arXiv preprint arXiv:2404.18852},
  year = {2024},
  url = {https://arxiv.org/abs/2404.18852}
}

@article{munteanu2018strongly,
  title={Strongly connected components-Algorithm for finding the strongly connected components of a graph},
  author={Munteanu, Vlad-Andrei},
  journal={arXiv preprint arXiv:1802.05387},
  year={2018},
  doi = {
https://doi.org/10.48550/arXiv.1802.05387
}
}

@manual{oracle-javac,
  title        = {javac: The Java Compiler},
  organization = {Oracle},
  year         = {2026},
  url          = {https://docs.oracle.com/en/java/javase/17/docs/specs/man/javac.html},
  note         = {Accessed: 2026-04-09}
}

@misc{oracle-classnotfound,
  title        = {ClassNotFoundException (Java SE API)},
  author       = {{Oracle}},
  year         = {2026},
  url          = {https://docs.oracle.com/en/java/javase/17/docs/api/java.base/java/lang/ClassNotFoundException.html},
  note         = {Accessed: 2026-04-09}
}

@misc{commons-cli,
  author       = {{Apache Commons CLI Project}},
  title        = {Apache Commons CLI},
  year         = {2026},
  howpublished = {\url{https://github.com/apache/commons-cli}},
  note         = {Accessed: 2026-04-13}
}

@misc{commons-codec,
  author       = {{Apache Commons Codec Project}},
  title        = {Apache Commons Codec},
  year         = {2026},
  howpublished = {\url{https://github.com/apache/commons-codec}},
  note         = {Accessed: 2026-04-13}
}

@misc{commons-csv,
  author       = {{Apache Commons CSV Project}},
  title        = {Apache Commons CSV},
  year         = {2026},
  howpublished = {\url{https://github.com/apache/commons-csv}},
  note         = {Accessed: 2026-04-13}
}

@misc{commons-exec,
  author       = {{Apache Commons Exec Project}},
  title        = {Apache Commons Exec},
  year         = {2026},
  howpublished = {\url{https://github.com/apache/commons-exec}},
  note         = {Accessed: 2026-04-13}
}

@misc{commons-fileupload,
  author       = {{Apache Commons FileUpload Project}},
  title        = {Apache Commons FileUpload},
  year         = {2026},
  howpublished = {\url{https://github.com/apache/commons-fileupload}},
  note         = {Accessed: 2026-04-13}
}

@misc{commons-text,
  author       = {{Apache Commons Text Project}},
  title        = {Apache Commons Text},
  year         = {2026},
  howpublished = {\url{https://github.com/apache/commons-text}},
  note         = {Accessed: 2026-04-13}
}

@misc{guardclauses,
  author       = {{Ardalis}},
  title        = {GuardClauses: A Simple Package with Guard Clause Extensions},
  year         = {2026},
  howpublished = {\url{https://github.com/ardalis/GuardClauses}},
  note         = {Accessed: 2026-04-13}
}

@misc{mediatr,
  author       = {{Lucky Penny Software}},
  title        = {MediatR: Simple Mediator Implementation in .NET},
  year         = {2026},
  howpublished = {\url{https://github.com/LuckyPennySoftware/MediatR}},
  note         = {Accessed: 2026-04-13}
}

@misc{csharp-mcp,
  author       = {{InfinityFlowApp}},
  title        = {csharp-mcp: C\# MCP Framework},
  year         = {2026},
  howpublished = {\url{https://github.com/InfinityFlowApp/csharp-mcp}},
  note         = {Accessed: 2026-04-13}
}

@misc{maven-surefire-errors,
  title        = {Maven Surefire Plugin: Error Summary},
  author       = {{Apache Maven Project}},
  year         = {2026},
  url          = {https://maven.apache.org/surefire/maven-surefire-plugin/newerrorsummary.html},
  note         = {Accessed: 2026-04-09}
}

@article{Nitin2024SpecTra,
  author = {Nitin, Vikram and Krishna, Rahul and Ray, Baishakhi},
  title = {SpecTra: Enhancing the Code Translation Ability of Language Models by Generating Multi-Modal Specifications},
  journal = {arXiv preprint arXiv:2405.18574},
  year = {2024},
  url = {https://arxiv.org/abs/2405.18574}
}

@online{meta2024java2kotlin,
  author    = {Jocelyn Luizzi and Jingbo Yang and Eve Matthaey},
  title     = {Translating Java to Kotlin at Scale},
  year      = {2024},
  month     = dec,
  day       = {18},
  url       = {https://engineering.fb.com/2024/12/18/android/translating-java-to-kotlin-at-scale/},
  note      = {Engineering at Meta},
}

@online{jetbrains2026kotlinvscode,
  author    = {{JetBrains}},
  title     = {Java to Kotlin Conversion Comes to Visual Studio Code},
  year      = {2026},
  month     = feb,
  url       = {https://blog.jetbrains.com/kotlin/2026/02/java-to-kotlin-conversion-comes-to-visual-studio-code/},
  note      = {Accessed: 2026-04-07}
}

@online{android2024kotlinfirst,
  author    = {{Android Developers}},
  title     = {Android's Kotlin-first approach},
  year      = {2024},
  url       = {https://developer.android.com/kotlin/first},
  note      = {Accessed: 2026-04-07}
}

@ARTICLE{RepoTransBench2024,
  author={Wang, Yanli and Wang, Yanlin and Wang, Suiquan and Guo, Daya and Chen, Jiachi and Grundy, John and Liu, Xilin and Ma, Yuchi and Mao, Mingzhi and Zhang, Hongyu and Zheng, Zibin},
  journal={IEEE Transactions on Software Engineering}, 
  title={RepoTransBench: A Real-World Multilingual Benchmark for Repository-Level Code Translation}, 
  year={2026},
  volume={52},
  number={2},
  pages={675-690},
  doi={10.1109/TSE.2025.3645056}}

@ARTICLE{KotlinMigrationStudy,
author={Martinez, Matias and Gois Mateus, Bruno},
journal={ IEEE Transactions on Software Engineering },
title={{ Why Did Developers Migrate Android Applications From Java to Kotlin? }},
year={2022},
volume={48},
number={11},
ISSN={1939-3520},
pages={4521-4534},
doi={10.1109/TSE.2021.3120367},
url = {https://doi.ieeecomputersociety.org/10.1109/TSE.2021.3120367},
publisher={IEEE Computer Society},
address={Los Alamitos, CA, USA},
month=nov}

@inproceedings{KotlinJavaDeps,
author = {Feng, Qiong and Ji, Huan and Ma, Xiaotian and Liang, Peng},
title = {Cross-Language Dependencies: An Empirical Study of Kotlin-Java},
year = {2024},
isbn = {9798400710476},
publisher = {Association for Computing Machinery},
address = {New York, NY, USA},
url = {https://doi.org/10.1145/3674805.3686680},
doi = {10.1145/3674805.3686680},
booktitle = {Proceedings of the 18th ACM/IEEE International Symposium on Empirical Software Engineering and Measurement},
pages = {189–199},
numpages = {11},
location = {Barcelona, Spain},
series = {ESEM '24}
}

@article{MigrationExp,
title = {Learning migration models for supporting incremental language migrations of software applications},
journal = {Information and Software Technology},
volume = {153},
pages = {107082},
year = {2023},
issn = {0950-5849},
doi = {https://doi.org/10.1016/j.infsof.2022.107082},
url = {https://www.sciencedirect.com/science/article/pii/S0950584922001914},
author = {Bruno Góis Mateus and Matias Martinez and Christophe Kolski}
}

@inproceedings{pan2024lost,
  author = {Pan, Rangeet and Ibrahimzada, Ali Reza and Krishna, Rahul and Sankar, Divya and Wassi, Lambert Pouguem and Merler, Michele and Sobolev, Boris and Pavuluri, Raju and Sinha, Saurabh and Jabbarvand, Reyhaneh},
  title = {Lost in Translation: A Study of Bugs Introduced by {Large Language Models} while Translating Code},
  year = {2024},
  booktitle = {ICSE '24: Proceedings of the IEEE/ACM 46th International Conference on Software Engineering},
  articleno = {82},
  numpages = {13},
  location = {Lisbon, Portugal},
  doi = {10.1145/3597503.3639226}
}

@inproceedings{hilton,
author = {Hilton, Michael and Bell, Jonathan and Marinov, Darko},
title = {A large-scale study of test coverage evolution},
year = {2018},
booktitle = {ASE '18: Proceedings of the 33rd ACM/IEEE International Conference on Automated Software Engineering},
isbn = {9781450359375},
publisher = {Association for Computing Machinery},
address = {New York, NY, USA},
url = {https://doi.org/10.1145/3238147.3238183},
doi = {10.1145/3238147.3238183},
pages = {53–63},
numpages = {11},
location = {Montpellier, France},
series = {ASE '18}
}

@inproceedings{yang2023coderepresentation,
  author={Yang, Lin and Chen, Junjie and You, Hanmo and Han, Jiachen and Jiang, Jiajun and Sun, Zhe and Lin, Xinqi and Liang, Fang and Kang, Yuning},
  booktitle={2023 IEEE 34th International Symposium on Software Reliability Engineering (ISSRE)}, 
  title={Can Code Representation Boost IR-Based Test Case Prioritization?}, 
  year={2023},
  volume={},
  number={},
  pages={240-251},
  doi={10.1109/ISSRE59848.2023.00077}}


\end{document}